\documentclass[preprint2]{aastex}

\usepackage{color, rotate}
\shorttitle{SN Ib/c Progenitors in Binary Systems}
\shortauthors{Kim, Yoon, \& Koo}

\begin{document}
\title{Observational Properties of Type Ib/c Supernova Progenitors 
in Binary Systems}

\author{Hyun-Jeong Kim\altaffilmark{1},
Sung-Chul Yoon\altaffilmark{2}, 
and Bon-Chul Koo\altaffilmark{3}}
\affil{Department of Physics and Astronomy, 
Seoul National University, Gwanak-ro 1, Gwanak-gu, \\ 
Seoul 151-742, Korea}
\altaffiltext{1}{hjkim@astro.snu.ac.kr}
\altaffiltext{2}{yoon@astro.snu.ac.kr }
\altaffiltext{3}{koo@astro.snu.ac.kr}
\begin{abstract}
In several recent observational studies on Type Ib/c supernovae (SNe Ib/c), the
inferred ejecta masses have a peak value of 2.0 -- 4.0~$M_\sun$, in favor
of the binary scenario for their progenitors rather than the Wolf-Rayet star
scenario.  To investigate the observational properties of relatively low-mass
helium stars in binary systems as SN Ib/c progenitors, we constructed
atmospheric models with the non-LTE radiative transfer code CMFGEN, using
binary star evolution models. We find that these helium stars can be
characterized by relatively narrow helium emission lines if the mass-loss rate
during the final evolutionary phase is significantly enhanced 
as implied by many SN Ib/c observations.
The optical brightness of helium star progenitors can be meaningfully
enhanced with a strong wind for $M \gtrsim 4.4~M_\sun$, 
but hardly affected or slightly weakened for relatively
low-mass of $\sim 3.0~M_\odot$, compared to the simple estimate using blackbody approximation.
We further confirm the previous suggestion
that the optical brightness would be generally higher for a less massive SN
Ib/c progenitor.  In good agreement with previous studies, our results indicate
that the optical magnitudes and colors of the recently detected progenitor of
the SN Ib iPTF13bvn can be well explained by a binary progenitor with a final
helium star mass of about 3.0 -- 4.4~$M_\sun$.  
\end{abstract}

\section{Introduction}

It is well known that the majority of massive stars are born in
multiple systems, and
binary systems play  an important role for core-collapse supernovae (SNe). 
Recent observations indicate that a large fraction of
massive binary stars are in relatively short-period orbits, implying
that they would undergo binary interactions during the course of their
evolution~\citep{sana12}.  Furthermore, the mass ratios of the stellar
components of massive binary systems are found to be close to one for many
cases, providing a favorable condition for stable mass transfer to avoid
mergers~\citep{kobulnicky07}.  These findings strengthen the long-standing
argument that binary interaction should be considered one of the primary
factors for massive star evolution, and that binary stars may not only be
related to certain exotic SNe, but also to commonly observed ones including
Type Ib/c supernovae (SNe Ib/c) \citep[e.g.,][]{podsiadlowski92, woosley95, wellstein99, eldridge08,
yoon10, smith11} .

SNe Ib/c, which constitute about 25\% of all core-collapse SNe~\citep{smith11,
eldridge13}, are therefore important to constrain the evolution of massive
stars.  The hydrogen envelopes of the progenitors of these SNe must have been
stripped off before they exploded.  Both mass loss from single stars and binary
interactions can fulfill this condition, but the resulting properties of SN
Ib/c progenitors would be very different from each other~\citep{yoon12,
eldridge13}.  Single Wolf-Rayet (WR) stars in the nearby Universe have
bolometric luminosities of $\log L/L_\odot \gtrsim 5.1$, implying that their
masses are higher than about $10~M_\odot$.  Their final masses at the pre-SN
stage are predicted to be higher than about $7~M_\odot$.  By contrast, lower
final masses are strongly preferred in binary progenitors.  Recent analyses on
the light curves of SNe Ib/c indicate that SN Ib/c ejecta masses are typically
around 1.0--5.0~$M_\odot$~\citep{drout11, cano13, lyman14, taddia15},
supporting the binary scenario over the WR star scenario.  

However, the debate on which type of progenitors between single WR stars and
relatively low-mass helium stars in binary systems is the dominant one for SNe
Ib/c is still on-going \citep[see][for a recent review]{yoon15}.  
The best way to resolve the issue would be therefore to directly identify SN
Ib/c progenitors in pre-SN images~\citep{smartt09}. The previous
searches have not been successful~\citep{crockett07, smartt09, eldridge13},
except for the tentative identification of the progenitor of the SN Ib
iPTF13bvn~\citep{cao13}.  

\citet{yoon12} made rough estimates on the optical magnitudes of SN Ib/c
progenitors at the pre-SN stage using their evolutionary models of
massive stars. They concluded that WR progenitors would be generally very faint
in the optical ($M_V \approx -2 \sim -3$ mag) compared to most WR stars observed in
the nearby Universe, while relatively low-mass helium star progenitors in binary
systems can be much brighter in the optical, as they become helium giant stars.
\citet{eldridge15} also made a similar conclusion on binary progenitors.  But
this conclusion is based on the simple assumption of blackbody radiation from
the adopted stellar evolution models.  In reality,  both absorption and
emission lines from helium stars may have a significant impact on the optical
brightness, depending on the surface temperature and the mass-loss rate by
winds. 

\citet{groh13b} and \citet{groh13a} presented stellar atmospheric models of
single WR type progenitors at the pre-SN stage using the stellar
evolution models of the Geneva group. They concluded that WO type progenitors,
of which the initial masses are higher than about 35~$M_\sun$, would be
relatively faint in the optical, in good agreement with \citet{yoon12}, and
that  WN type progenitors, which are expected for a limited initial mass range
($M = 31-35~M_\odot$; but see \citealt{eldridge15}), can have optical magnitudes of
 about $-5.5 \sim -6.5$ mag.  However, there exists no such a study on binary
progenitors.  

The purpose of this paper is therefore to present stellar atmospheric models of
binary progenitors of SNe Ib/c, for the first time.  This would allow to better
test the binary progenitor scenario with iPTF13bvn as well as future
observations.  In Sect.~\ref{sect:assumptions}, we explain our numerical
methods and the adopted physical assumptions. The result of our
calculations is summarized in Sect.~\ref{sect:results}. We compare our model
spectra with those of HD 45166 and $\upsilon$ Sgr which are relatively low-mass
helium stars observed in our Galaxy, as well as WN stars in
Sect.~\ref{sect:counterpart}. The predicted optical magnitudes and the
implications for the progenitor of iPTF13bvn are discussed in
Sects.~\ref{sect:magnitudes} and~\ref{sect:iptf}.  We conclude this study in
Sect.~\ref{sect:conclusions}.

\section{Physical assumptions}~\label{sect:assumptions}

The helium star progenitor models for the present study were chosen from the
binary evolution models by \citet{yoon10}.  These models were calculated at
solar metallicity including the effects of rotation, and followed up to the end
of core neon burning, which is about one year before core collapse.  The
surface properties of these models remain almost unchanged during the final
evolutionary stages after carbon exhaustion, which last for about 100 years.
The pre-SN images used for the searches of SN Ib/c progenitors were usually
taken about 10 years before the SN explosion~\citep[e.g.,][]{crockett07, cao13}
and therefore any model after core carbon exhaustion are suitable for
comparison with observations.  The models chosen for atmospheric calculations
are the last computed models of the sequences 5, 22 and 27, which have the
final masses of 3.0, 4.4, and 5.1~$M_\odot$, respectively.  The initial masses
of these models are 14, 18, and 25~$M_{\sun}$, respectively.  The optical
luminosity of a helium star at the pre-SN stage is a sensitive function of its
mass and radius~\citep{yoon12}.  \citet{yoon10} showed that there exists a
fairly good mass-radius relation for binary SN Ib/c progenitors at the pre-SN
stage, and therefore these selected models can roughly represent    the surface
properties of SN Ib/c progenitors for the given final masses.

Given the neutron star remnant mass of about 1.3~$M_\odot$, this range of
helium star masses is consistent with the peak values of the SN Ib/c ejecta
mass distribution (i.e., $M_\mathrm{ej} \approx 2 - 4~M_\odot$) inferred from
SN light curves~\citep{drout11, cao13, lyman14}.  It also fits well with the
inferred ejecta mass of iPTF13bvn ($M_\mathrm{ej} \approx 2.0~M_\odot$;
\citealt{fremling14, bersten14}), and thus we can directly compare our result
to the observed properties of the tentative iPTF13bvn progenitor.

\citet{yoon10} adopted the WR mass-loss rate given by \citet{hamann95} using a
reduction factor of 5 - 10 to consider the effect of WR wind clumping. However,
the resulting wind mass-loss rates of these models are not high enough to have
an optically thick WR type wind, in particular for 3.0 and 4.4~$M_\sun$ helium
stars, given that their masses are relatively low and that their envelopes have
relatively large radii.  But the mass-loss rates from helium stars of the
considered mass range are not well constrained by observations.  These helium
stars undergo very rapid increase in the surface luminosity as they approach to
the pre-SN stage \citep{yoon10, yoon12}, and we cannot exclude the possibility
that mass loss becomes dramatically strong during this final stage (i.e.,
$\dot{M} \gtrsim 10^{-5}~M_\sun~\mathrm{yr^{-1}}$), which is in fact implied by
many SN Ib/c observations~\citep[e.g.,][]{Foley07, Pastorello08, Wellons12,
Gorbikov14}. Therefore, we also consider an arbitrarily enhanced mass-loss rate
for each helium star model as summarized in Table~\ref{tbl1}.  Here, the label
w1 denotes the models with the mass-loss rates used by \citet{yoon10}, and w2
the models with the enhanced mass-loss rates.

The spectra of our SN Ib/c progenitor models were computed using the non-local
thermodynamical equilibrium atmospheric radiative transfer code CMFGEN
\citep{hillier98, hillier03}.  CMFGEN determines the temperature
distribution of the expanding atmosphere by solving the statistical and
radiative equilibrium equations, and  computes line and continuum formation
with the spherical symmetric geometry \citep{hillier90}.  CMFGEN uses the
super-level approach to fully treat line blanketing. In this approach, levels
with similar excitation energies are grouped into a single level, under the
assumption that the departure coefficients in a group are identical, and only
the population of the super level is solved to to specify the populations of
the levels within a super level \citep{hillier98,hillier03}.  

CMFGEN requires a complete (previously converged) model 
including atmospheric structure, atomic models and
their departure coefficients as the initial trial solution.
For our calculations, we first adopted one of the pre-calculated O star 
models provided by Hillier\footnote{http://kookaburra.phyast.pitt.edu/hillier/web/CMFGEN.htm}
with the effective temperature of 27,500~K and ${\rm Log}~g$ of 3.25,  
which are similar to those of our 4.4~$M_\sun$ helium star (Table~\ref{tbl1}). 
Table~\ref{tbl2} presents the atomic species included in our calculations 
with the numbers of super and full levels. The atomic data and information
on the levels are accompanied with the CMFGEN code.
Once we obtained the converged model for 4.4~$M_{\sun}$ helium stars, 
we used it as the initial trial solution for the other models. 
The atomic models from the above O star do not include 
neutral and low-excitation level 
metal lines, which may be relevant for our 3.0~$M_{\sun}$ model, 
but this does not significantly affect the main conclusions of our study
as explained in  Section~\ref{sect:counterpart} below.

While a hydrostatic solution for the subsonic part is self-consistently solved
by CMFGEN, it does not solve the momentum equation of the wind, and thus the
wind velocity profile needs to be specified.  For the wind part, we assume the
standard $\beta$-type velocity law.  We used $\beta=1$ and 1.5 and
$v_\infty/v_\mathrm{esc} =$ 2.0 and 1.5 for the optically thin wind models (w1)
and the WR-type wind models (w2), respectively \citep[cf.][]{vink05}, where
$v_\infty$ and $v_\mathrm{esc}$ respectively denote the terminal wind velocity
and the escape velocity.  The velocity structure is modified at depth to
smoothly match the structure at the surface of the hydrostatic core \citep[for
details on the CMFGEN calculations, see][]{hillier90,hillier98,hillier03}.

\citet{yoon10} considered massive companion stars of OB type in their
calculation.  The companion star masses of the chosen progenitor models ($M =
3.0, 4.4, ~\mathrm{and}~ 5.1~M_\sun$) at the end of the calculation are 18, 23,
and 35~$M_\sun$, respectively.  Note that these are not unique solutions. 
\citet{yoon10} followed the spin-up of mass accreting star as a result of mass and
angular momentum accretion and the resulting enhancement of mass loss, and
thus the mass transfer process is highly non-conservative in their
calculations.  However, the companion star masses can significantly vary
according to the adopted  mass accretion efficiency (i.e., the ratio of the
transferred mass to the accreted mass).  For example, if the mass transfer were
conservative, a binary system where both stars have an initial mass of about
20~$M_\sun$ could produce a  3 $M_\sun$ helium star that has a structure
similar to that of the 3.0~$M_\sun$ models of the present study, via Case A and AB
mass transfers.  The corresponding companion star mass would be about
35~$M_\sun$  in this case \citep[cf.][]{wellstein99}.  

Given that many different combinations of helium star and companion star masses
are possible in principle, we included  O-type star models of three different masses
($M=$~20, 30, and 35~$M_{\sun}$) in the CMFGEN computation to investigate how a
luminous O-type star companion would influence the optical brightness of SN Ib/c
progenitor systems. These masses were chosen mainly because they can give the
best fit with the observational properties of the iPTF13bvn progenitor
candidate as discussed below.   The 20 and 30~$M_{\sun}$  models are
non-rotating models without overshooting in their early stages of the main
sequence that are close to the zero-age main sequence.  The 35~$M_{\sun}$ model
is the last computed companion star model of the sequence  27  of
\citet{yoon10} (i.e, the sequence for the 5.1~$M_\sun$ helium star model in the
present study). 

The model parameters are summarized in Table~\ref{tbl1}.  The table lists two
kinds of temperature and radius. The values of $T_{\star}$ and $R_{\star}$ are
predicted from the stellar evolution code \citep{yoon10} without correcting the
optical depth effects from the wind, whereas the effective temperature ($T_{\rm
eff}$) and photospheric radius ($R_{\rm phot}$) are the outputs from the CMFGEN
calculations defined at the layer where the Rosseland optical depth ($\tau_{\rm
ross}$) is 2/3.  In the calculations, hydrodynamic clumping in the wind is
considered by adopting a volume filling factor ($f$). For all models, we assume
$f=0.1$.  The surface abundances were taken from the data of the selected
evolutionary models.

\section{Results of the atmospheric models}\label{sect:results}

\begin{figure}[t]
\includegraphics[width=\columnwidth]{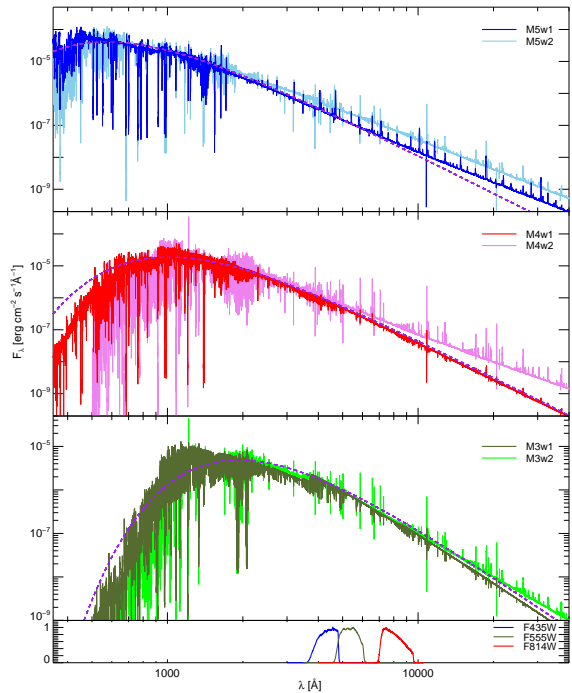}
\caption{SEDs of the helium star models 
and the blackbody fluxes (purple dashed line) of 
the given temperature $T_{\star}$ and luminosity 
of each mass model. 
All fluxes are scaled to a distance of 10~pc. Transmission 
curves of the {\it HST}/ACS Wide Field Channel filters, {\it F435W}, {\it F555W}, 
and {\it F814W} are also presented at the bottom panel.
\label{fig1}}
\end{figure}

Spectral energy distributions (SEDs) of the helium stars from the CMFGEN
calculations are shown in Figure~\ref{fig1}.  All fluxes are scaled to a
distance of 10~pc.  The SEDs of the w1 and w2 cases for a given helium star
mass are almost similar to each other, but the w2 models have excess in long
wavelengths because of the extended photosphere radius with an optically thick
wind.  For comparison, we also present the blackbody fluxes of the given
temperature ($T_{\star}$) and luminosity for each model.  In optical wavelength
ranges, the differences between the blackbody fluxes and the helium star models
are not significant.  However, the w2 models, particularly for $M =  4.4$ and
5.1~$M_{\sun}$, show strong excess in continuum as well as a wealth of emission
lines in the optical.  The large differences in the extreme ultraviolet
wavelengths are owing to the line blanketing effect. 

\begin{figure}[t]
\includegraphics[width=\columnwidth]{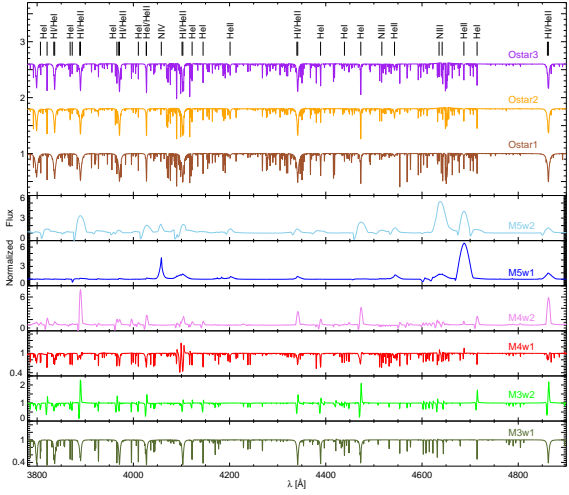}
\caption{Normalized optical spectra of the helium star 
and O-type star models for the wavelength range of $3750~\mathrm{\AA} \le \lambda \le 4900~\mathrm{\AA}$.}
\label{fig2}
\end{figure}

\begin{figure}[t]
\includegraphics[width=\columnwidth]{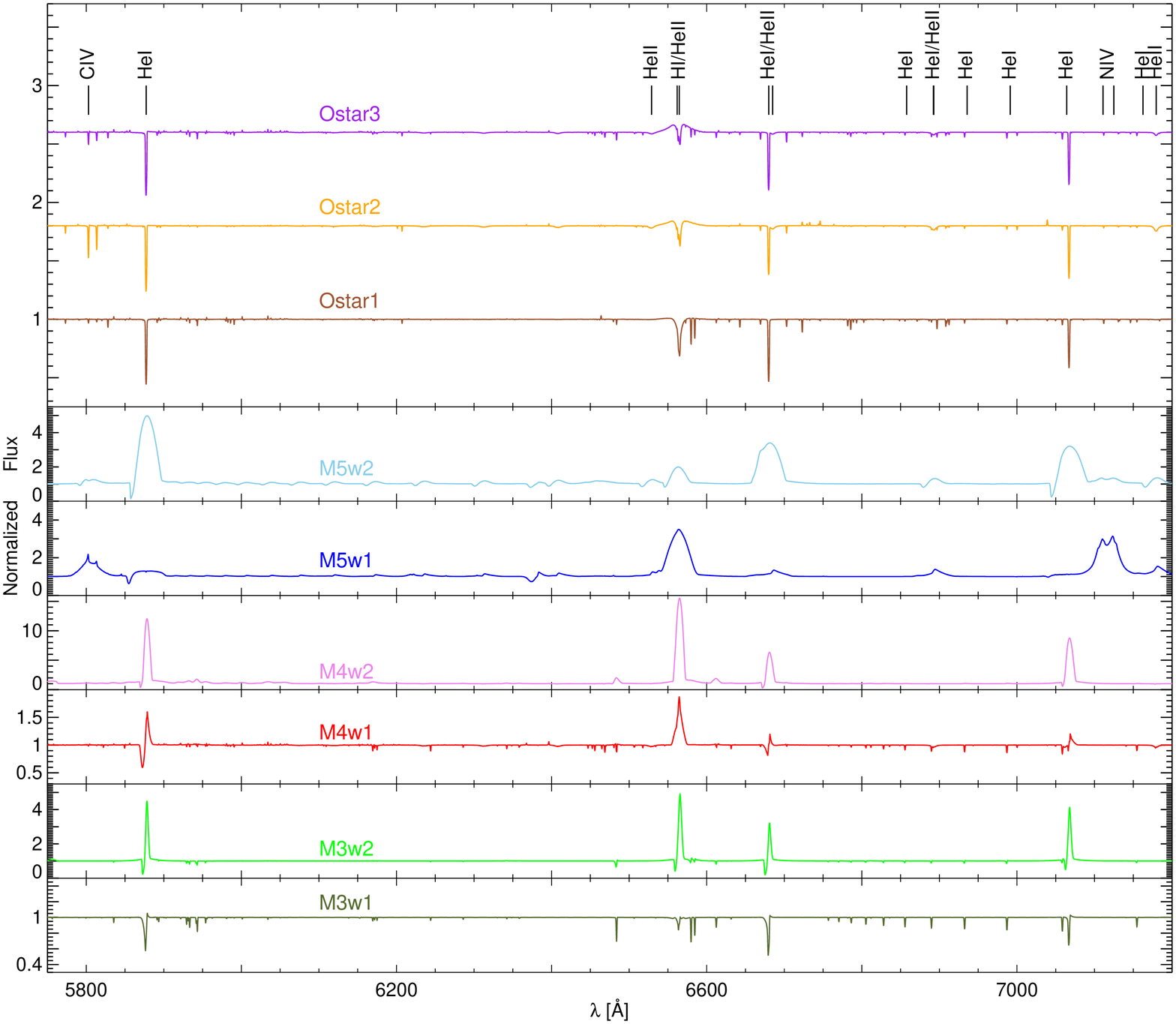}
\caption{Normalized optical spectra of the helium star 
and O-type star models for the wavelength range of $5750~\mathrm{\AA} \le \lambda \le 7200~\mathrm{\AA}$.}
\label{fig3}
\end{figure}

Figures~\ref{fig2}~and~\ref{fig3} present the normalized optical spectra of all models. While
the spectra of O-type stars only show absorption lines, those of helium stars
(particularly the w2 models) show emission lines in general.  The
3.0~$M_{\sun}$ models and the 4.4~$M_{\sun}$ model with w1 type wind (M3w1,
M3w2, and M4w1) dominantly show absorption lines; however, compared to O-type stars,
He~\textsc{i} lines of the helium stars are broader with developed wings (e.g.,
He~\textsc{i}~$\lambda$4024 or He~\textsc{i}~$\lambda$4472), and they sometimes
appear as (weak) P Cygni profiles (e.g., He~\textsc{i}~$\lambda$6679 or
He~\textsc{i}~$\lambda$7067).  In addition, the H$\alpha$ line of the M4w1 model
appears in emission\footnote{ As discussed by \citet{yoon10} in detail,  the
surface mass fraction of hydrogen for the 3.0 and 4.4~$M_\sun$ models is
about 0.01 (Table~\ref{tbl1}).  Hydrogen lines are actually detected in many
SNe classified as Type Ib (e.g., \citealt{deng00, branch02, elmhamdi06,
spencer10}), and this theoretical prediction of the tiny amounts of hydrogen in
some SN Ib/c progenitors  is consistent with observations~\citep[see
also][]{dessart11}.}.  These emission lines make the helium star to be
distinguishable from O stars with a similar temperature and mass-loss rate.
The H lines in the M3w1 model are absorption lines and weaker than those of
O-type stars because of the lower abundance of hydrogen (see Table~\ref{tbl1}).
The spectra of the 5.1~$M_{\sun}$ models of which hydrogen is almost depleted
show broad helium emission lines and high-ionization lines such as
N~\textsc{iii}/\textsc{iv}. 

\section{Comparison with Observational Counterparts}\label{sect:counterpart}

One of the observational counterparts of binary SN Ib/c progenitors is the
quasi-WR (qWR) star HD 45166, which is composed of  a helium rich
4.2~$M_{\sun}$ star with $R \simeq 1.0~R_{\sun}$ and a 4.8~$M_{\sun}$ main
sequence star (B7V) in a 1.596 day orbit \citep{willis83,steiner05,groh08}.
The optical spectrum of the helium star that is likely on the helium main
sequence shows a number of emission lines such as He~\textsc{i}/\textsc{ii},
N~\textsc{iii}/\textsc{iv}/\textsc{v}, and C~\textsc{iii}/\textsc{iv}, and it
is well explained by a model with $T_{\rm eff} = 50,000\pm2,000$~K
($T_{\star}=70,000\pm20,000$~K), ${\rm log}(L/L_{\sun})=3.75\pm0.08$, and
$\dot{M} = 2.2 \times 10^{-7}~M_{\sun}~{\rm yr}^{-1}$ \citep{groh08}.  The
spectrum of the qWR star in HD 45166 with weak He~\textsc{i} lines and
higher-ionization lines (e.g., N~\textsc{iv} or C~\textsc{iv}) is different
from our helium star model spectra of similar mass (M4w1 and M4w2), which
dominantly show He~\textsc{i} lines.  This difference is likely due to the much
lower effective temperature ($T_{\rm eff} =$ 20,000--30,000~K) of our
4.4~$M_{\sun}$ models.  As discussed in \citet{yoon10,yoon12}, low-mass ($M
=$~3--5~$M_{\sun}$) helium stars are hot and visually very faint on the helium
main sequence but become cooler and more luminous during the final evolutionary
stages because of the rapidly expanding envelopes.  The qWR star in HD 45166 is
in fact much fainter in the optical \citep[$M_{V}=-0.21$~mag;][]{willis83} than
our 4.4~$M_{\sun}$ models ($M_{V} \approx -5$~mag, see Section~3).  

\begin{figure}[th]
\includegraphics[width=\columnwidth]{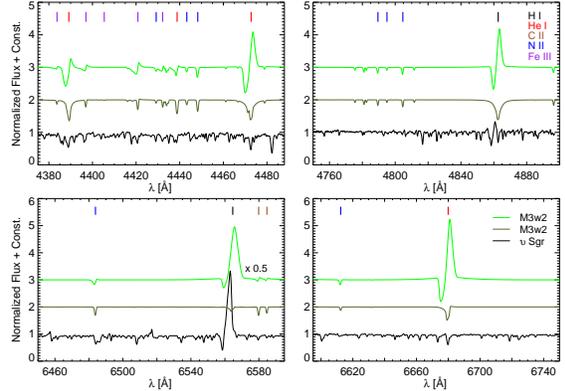}
\caption{Comparison of the $\upsilon$ Sgr spectrum (black) that was obtained 
by a 60-centimetre telescope at the Observatoire du Pic du Midi on July 24, 2008, 
which is available from the BeSS
database (http://basebe.obspm.fr/basebe),
with the 3.0~$M_{\sun}$ helium star model spectra (dark green and light green 
for M3w1 and M3w2, respectively).  Vertical lines at the top of each panel 
mark spectral lines seen in the both $\upsilon$ Sgr and helium star models. 
Black, red, brown, blue, and purple lines correspond to H~\textsc{i}, He~\textsc{i},
C~\textsc{ii}, N~\textsc{ii}, and Fe~\textsc{iii}, respectively. In the lower left 
panel, the M3w2 spectrum is scaled to 50\% for display.
\label{fig4}}
\end{figure}

Another observational counterparts are evolved helium giant stars
in binary systems. Only four stars of such a system are currently known
including $\upsilon$ Sgr, KS Per, LSS 1922, and LSS 4300 \citep{dudley93}, and
the $\upsilon$ Sgr system among them has been best studied. The primary star of
the $\upsilon$ Sgr system is a hydrogen-deficient star with $M \sim
3.0~M_{\sun}$, $T_{\rm eff} \sim 12,000$~K, and ${\rm log}L/L_{\sun} \simeq
4.6$ \citep{saio95,kipper12}, which are comparable to those of our 3.0~$M_\sun$
helium star models.  The main characteristics of the $\upsilon$ Sgr spectrum
includes He~\textsc{i} absorption lines, a large number of absorption/emission
lines of neutral and ionized metals from low excitation levels, some P Cygni
profiles, and the forbidden lines of low-ionization metal lines
(Figure~\ref{fig4}; \citealt{kipper12}).  He~\textsc{i} absorption lines and
metallic lines such as C~\textsc{ii}, N~\textsc{ii}, or Fe~\textsc{iii} present
similar characteristics for the $\upsilon$ Sgr spectrum and the M3w1 model
spectrum.  \citet{bonneau11} argue that the hydrogen P Cygni profiles of the
$\upsilon$ Sgr originate either from the circumbinary disk of this system or
from the disk around the unseen secondary star.  This implies that the wind
mass-loss rate from the $\upsilon$ Sgr is  comparable to the value adopted for
M3w1 model that do not show emission lines, rather than that of M3w2 model for
which the spectrum is characterized by strong emission lines for both hydrogen
and helium~(Figures~\ref{fig2} and~\ref{fig3}).  A number of other absorption
lines only seen in the $\upsilon$ Sgr spectrum are neutral and metal lines of
low excitation levels (e.g., N~\textsc{i}, Ne~\textsc{i}, or Fe~\textsc{ii}),
which are not included in our models.


Finally, the spectrum of the M5w2 model may be compared with the spectra of WN
stars with similar temperatures. 
In the M5w2 model spectrum (Figures~\ref{fig2} and \ref{fig3}), helium and
nitrogen lines are dominantly observed, and most helium lines are broad and
appear as P Cygni profiles.  These spectral characteristics can also be found
in the spectra of WN stars \citep{hamann06}. In particular, WN8 stars with
$T_{\star} \simeq $ 40,000-50,000~K show very similar spectra with strong
He~\textsc{i} lines and negligibly weak C~\textsc{iv} line (at 5803\AA) as seen
in the M5w2 model spectrum.  As moving toward earlier type WN stars (i.e.,
higher surface temperature), He~\textsc{i} lines become weaker and the
C~\textsc{iv}~$\lambda$5803 line becomes stronger.  Compared to WN8 star
spectra \citep[e.g., WR012, WR040, or WR170 in][]{hamann06}, however, in the
M5w2 model spectrum He~\textsc{i} lines tend to be stronger, and the
N~\textsc{iii}~$\lambda$4635/41 line is stronger than the
He~\textsc{ii}~$\lambda$4687 line.  In addition, in spite of similar
temperature, the M5w2 model is fainter in the optical ($M_{V} \gtrsim -5$~mag,
see Section~3) than WN8 stars \citep[$M_{V} \sim -7$~mag;][]{hamann06}, which
is a natural consequence of the lower bolometric luminosity for the given
surface temperature.  Compared to the case of M5w2, in the spectrum of M5w1 the
lines from higher excitation species such as N~\textsc{iv} or C~\textsc{iv} are
stronger and He~\textsc{i} lines almost disappear likely due to higher
effective temperature compared to the M5w2 model (Table~\ref{tbl1}).

\section{Absolute Magnitudes of Helium Star Binary Progenitors}\label{sect:magnitudes}

\begin{figure*}[t]
\includegraphics[width=\textwidth]{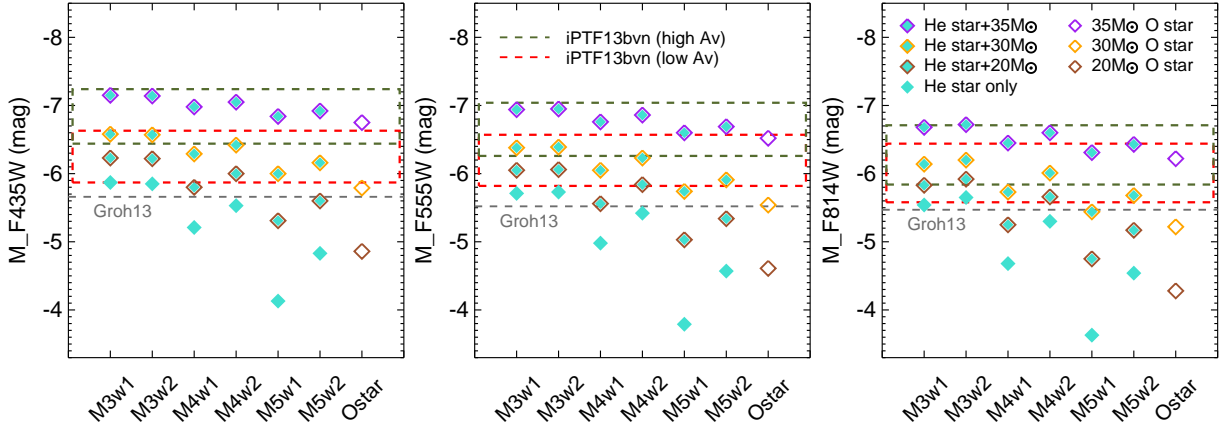}
\caption{Predicted absolute magnitudes of the models
in the {\it HST}/ACS {\it F435W (left)}, 
{\it F555W (middle)}, and {\it F814W (right)} filters.
Filled cyan diamonds without borders are
the absolute magnitudes of the helium star models for 
given {\it HST}/ACS filters, and open diamonds of brown, 
yellow, and purple are the magnitudes of 20, 30, and 
35~$M_{\sun}$ O-type stars, respectively. 
The filled cyan diamonds with borders are the predictions 
of the helium stars in binary systems with an O-type star 
companion of corresponding colors. 
The extinction-corrected magnitude ranges of 
the iPTF13bvn progenitor candidate 
by the low \citep[$E(B-V)=0.0715$,][]{cao13} and
high \citep[$E(B-V)=0.2147$,][]{bersten14} extinction 
values are presented with red and dark green dashed 
boxes, respectively.
Gray dashed lines present the prediction of 32~$M_{\sun}$ 
WR star progenitor model from \citet{groh13a}.
\label{fig5}}
\end{figure*}

In Table~\ref{tbl3} and Figure~\ref{fig5}, we present the predicted magnitudes
of the helium star models for the {\it HST}/ACS filters of {\it F435W}, {\it
F555W} and {\it F814W},  which roughly correspond to Johnson~$B$, Johnson~$V$,
and Broad~$I$ filters, respectively \citep{sirianni05}.  We computed the
absolute magnitudes of the models in the VEGAMAG system using the synthetic
model spectra and the transmission curves of the {\it HST}/ACS filters
presented in Figure~\ref{fig1}. For a filter $P$ with a transmission curve of
$P(\lambda)$, the absolute magnitudes in the VEGAMAG system are given by ${\rm
VEGAMAG}(P) =$\\ 
$-2.5~{\rm log_{10}} [\int P(\lambda)
F_{\lambda}(\lambda)\lambda~d\lambda/ \int P(\lambda) F_{\lambda,{\rm
Vega}}(\lambda)\lambda~d\lambda]$, where $F_{\lambda}$ and $F_{\lambda,{\rm
Vega}}$ are the model flux from the CMFGEN calculations and the flux of Vega
scaled to a distance of 10~pc \citep{sirianni05}.  The reference spectrum of
Vega is adopted from the SYNPHOT package distributed as a part of STSDAS%
\footnote{http://www.stsci.edu/institute/software\_hardware/stsdas/synphot}.
We computed the bolometric magnitudes ($M_{\rm bol}$) as well, assuming that
the solar $M_{\rm bol}$ is 4.74~mag \citep{cox00}, and present bolometric
corrections in a given filter $P$ (BC$_{P} = M_{\rm bol}-{\rm VEGAMAG}(P)$) in
Table~\ref{tbl3}.

For comparison, the magnitudes and bolometric corrections of the blackbody
models are also presented in Table~\ref{tbl3}.  The deviation from the
blackbody model prediction  depends on the strength of various lines, as well
as the location of the actual photosphere for a given  wind mass-loss rate.
For example, if the photosphere were significantly lifted up with an optically
thick wind that produces strong emission lines, the resulting luminosity in the
optical would be much higher than in the corresponding case of blackbody. 
Otherwise, numerous absorption lines tend to decrease the visual brightness
compared to what the blackbody models predict.

In general, for a given mass-loss rate, the optical thickness of a stellar wind
becomes larger for a smaller radius of the star~\citep[e.g.,][]{langer89}.
This explains why the difference between the stellar radius ($R_*$) and the
photospheric radius ($R_\mathrm{phot}$) becomes larger with a higher mass for a
given wind mass-loss rate as shown in Table~\ref{tbl3}.  This also tends to
make emission lines stronger.  Therefore, the visual brightness of helium stars
can be strongly influenced by winds, and the assumption of the blackbody may
significantly underestimate the brightness of a helium star, as shown with M4w2
and M5w2 models.  As also expected, for a given helium star mass, such
difference becomes larger with a higher mass-loss rate: the w2 model tends to
be brighter by up to $\sim$1~mag than the w1 model.  

On the other hand, M3w1, M3w2 and M4w1 models  have  numerous absorption lines
in their spectra, and their photosphere radii are not much different from the
stellar radii (see Section~\ref{sect:results} and Figures~\ref{fig2}
and~\ref{fig3}).  Their visual brightness is comparable or a little fainter
($\sim$0.3~mag) than the corresponding blackbody models, because of the
presence of absorption lines and the lack of emission lines.  The 3.0~$M_\sun$
model could be somewhat fainter than predicted in Table~\ref{tbl3} if we
included low-excitation species, which can produce more absorption lines in the
spectra (see the above discussion on $\upsilon$~Sgr in
Section~\ref{sect:counterpart}).  However, for more massive
models, our result would not be affected with the inclusion
of those species because the surface temperature is too high for the low-excitation
lines to make any significant impact~\citep[e.g.][]{gray09}. 

Note that the optical brightness becomes systematically lower for a higher
helium star mass, and that the considered helium stars in the present study are
predicted to be much brighter than massive single WR stars of WO type ($M_{\rm
ZAMS} \gtrsim 30 - 35 M_{\sun}$) at the pre-SN stage that would have $M_{BVI}
\simeq -3$~mag \citep{yoon12,groh13b}.  The deepest absolute {\it BVR}
magnitude limits of SN Ib/c progenitors in pre-explosion images to date are
between $-4$ and $-5$~mag \citep{crockett07,eldridge13}, and they are
comparable to the faintest one in our helium star models M5w1.  Our
work confirms the conclusion by \citet{yoon12} that the non-detection of SN
Ib/c progenitors does not necessarily imply binary progenitors rather than
massive WR progenitors.  To the contrary, binary progenitors are easier to be
found in the optical than massive WR progenitors of WO type that would be the
most common type of SN Ib/c progenitors from single stars~\citep{yoon12,
eldridge15}.  The probability of detecting binary SN Ib/c progenitors can
further increase with a luminous O-type star companion as shown in
Figure~\ref{fig5},  while a significant fraction of binary progenitors would
have a less luminous dwarf star or a compact object as a
companion~\citep{yoon15}.

\section{Comparison with the Progenitor Candidate of the Supernova iPTF13bvn}\label{sect:iptf}

\citet{cao13} have recently reported a tentative identification of the
progenitor of the SN Ib iPTF13bvn exploded in NGC 5806 (22.5~Mpc) from the
pre-explosion {\it HST}/ACS images.  The observed magnitudes of the object in
{\it HST}/ACS {\it F435W}, {\it F555W}, and {\it F814W} images range from 26 to
27 mag depending on the adopted photometry method \citep{cao13, eldridge15}.
The inferred absolute magnitudes range from -5.3 to -7.3 mag,  given the
uncertainties of the extinction towards NGC 5806.  \citet{cao13} measured the
Milky-Way (foreground) and the host galaxy reddening from the observed
Na~\textsc{i}~D lines as $E(B-V)_{\rm MW}=0.0278$ and $E(B-V)_{\rm
host}=0.0437$, whereas \citet{bersten14} adopted $E(B-V)_{\rm MW}=0.0447$
\citep{schlafly11} and derived $E(B-V)_{\rm host}=0.17\pm0.03$ using an
intrinsic-color law from a sample of observed SNe.  In this paper, we adopt the
photometry results of \citet{eldridge15} and consider both values of reddening
to compare the progenitor candidate of iPTF13bvn with  our progenitor models.  
In Figure~\ref{fig5}, we present the magnitude range of the
iPTF13bvn progenitor candidate dereddened by the low \citep[$E(B-V)=0.0715$
from][]{cao13} and high \citep[$E(B-V)=0.2147$ from][]{bersten14} extinction
values.

As shown in the figure, if we ignore the contribution from a companion star,
only the 3.0~$M_\sun$  progenitor can marginally satisfy the observed magnitude
range of the iPTF13bvn progenitor with the low-extinction.  For the
4.4~$M_\sun$ and 5.1~$M_\sun$ progenitor, the companion star should be more massive than
20~$M_\sun$ and 30~$M_\sun$, respectively. With a 35~$M_\sun$ companion, all
progenitor models give optical magnitudes compatible with those in the
high-extinction case, but they are a bit too bright to fit with the
low-extinction case.  Therefore, for the considered mass range of helium stars,
35~$M_\sun$ roughly gives the upper limit of the companion star mass.  From the
evolutionary point of view, a more massive helium star progenitor would
systematically have a more massive companion if the binary system underwent
stable mass transfers. For example, with stable Case B mass transfer systems
\footnote{Systems where mass transfer from the primary star starts during the
helium core contraction phase.}, the upper limit of the companion star mass
would be roughly about 27, 35 and 38~$M_\sun$, for 3.0, 4.4 and 5.1~$M_\sun$
helium star progenitors~\citep[see][for more details on the binary progenitor
evolution]{yoon15}. With Case AB mass transfer systems\footnote{Systems where
the first mass transfer starts during core hydrogen burning, followed by
another mass transfer phase during helium core contraction.}, this limit would
be higher by several solar masses~\citep[cf.][]{wellstein99}.  The lower
limit of the companion star mass is zero: a helium star progenitor of the
considered mass range would not have any companion if it were produced via Case
A mass transfer that leads to reversal of the SN order as discussed by
\citet{pols94}.

\begin{figure}[t]
\includegraphics[width=\columnwidth]{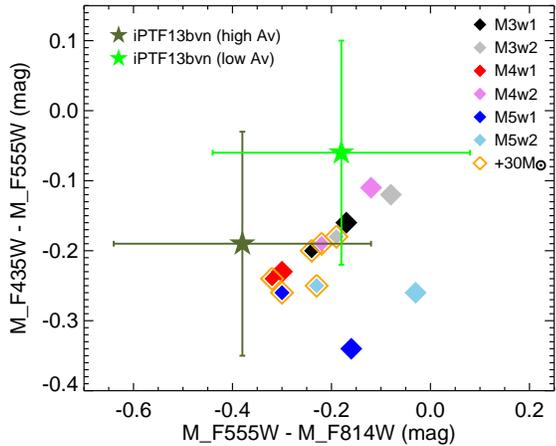}
\caption{Color-color diagram of ($F435W-F555W$) vs. 
($F555W-F814W$) of the helium star models 
without (filled diamonds) and with (filled diamonds with 
yellow borders) a 30~$M_{\sun}$ O-type star companion. 
Light and dark green filled stars are the colors of 
the iPTF13bvn progenitor candidate dereddened
by the low and high extinction values (see the caption
of Figure~\ref{fig5}), respectively.
\label{fig6}}
\end{figure}

Considering both the theoretical constraint and the rather large magnitude
uncertainty ($\sim$1~mag) resulting from the two extinction values, we conclude
that a helium star progenitor of 3.0/4.4~$M_{\sun}$ with a  20/30~$M_\sun$
O-type star companion  can best explain the observed magnitudes of the
progenitor candidate in all the three {\it HST}/ACS filters. Their
($F435W-F555W$) and ($F555W-F814W$) colors presented in Figure~\ref{fig6} are
also within the observed color ranges of the progenitor candidate, although all
of the helium star binary progenitor models are distributed in a narrow color
range. 

\citet{groh13a} suggested a single WR star of WN type with initial masses of
31--35~$M_{\sun}$ as a progenitor of iPTF13bvn based on the predicted
optical magnitudes ($M_{\rm V} \sim -5.5$ mag, gray dashed lines in
Figure~\ref{fig5}) of the non-rotating models from the Geneva stellar evolution
code.  We note that \citet{groh13a} used the photometry results from
\citet{cao13}, which are $\sim$1~mag fainter than those we present in this
study based on \citet{eldridge15}.  While the predicted optical brightness 
is within the error bar of the observation, the final mass of this model
($\sim$11~$M_{\sun}$) is too high to explain the typical ejecta masses of SNe
Ib/c \citep[$M_{\rm ejecta}=$ 1--5~$M_{\sun}$;][]{drout11,cano13, taddia15} as well as
the estimated ejecta mass of iPTF13bvn
\citep[$\sim$1.9--2.3~$M_{\sun}$;][]{bersten14,fremling14}. 

\begin{figure}[th]
\includegraphics[width=\columnwidth]{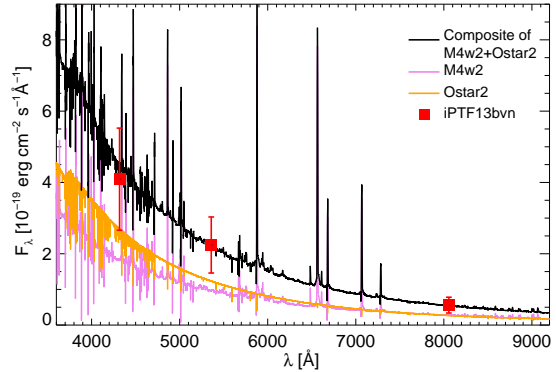}
\caption{Predicted spectrum of the binary progenitor model 
composed of a 4.4~$M_{\sun}$ helium star and a 30~$M_{\sun}$ O-type star companion (M4w2+Ostar2 model).
Pink and orange lines present the model spectra of 
the helium star and O-type star, respectively, and the black line 
is the composite spectrum of the two. The {\it HST}/ACS 
observations of the iPTF13bvn progenitor candiate 
\citep{eldridge15} dereddened by low-extinction value of 
$E(B-V)=0.0715$ \citep{cao13} are compared with red squares. 
\label{fig7}}
\end{figure}

Based on the SN ejecta mass and the optical brightness of the progenitor
candidate, a binary progenitor with an initial mass of 10--20~$M_{\sun}$ for
 iPTF13bvn has been suggested \citep{bersten14,eldridge15}. Our stellar
atmospheric modeling of low-mass helium stars ($M_{\rm final}=$
3--4~$M_{\sun}$) with an O-type star companion of $\sim$ 20 -- 30~$M_{\sun}$ also supports
this scenario (Figures~\ref{fig5} and \ref{fig6}).  \citet{bersten14} proposed
the binary system composed of a 3.7~$M_{\sun}$ helium star ($T_{\rm eff} \sim
16,000$~K) and a 33.7~$M_{\sun}$ hot companion ($T_{\rm eff} \sim 44,000$~K)
and predicted the SED of the progenitor at the pre-SN state.  Their model
\citep[Figure~5 of][]{bersten14} is fairly well consistent with the {\it
HST}/ACS observations, but the assumption of a blackbody for the helium
star may have uncertainty up to $\sim$1~mag as discussed in Secion~\ref{sect:magnitudes}. 
Figure~\ref{fig7} compares a predicted spectrum of one of our helium star
binary models (M4w2+Ostar2) that reproduce best the observations of the
iPTF13bvn progenitor candidate, which are dereddened by the low extinction
value ($E(B-V)=0.0715$).  In the composite spectrum of the progenitor model,
while the O-type star dominantly contributes the flux in the optical wavelengths,
emission lines expected from the helium star appear in the spectrum; this makes
the helium star binary system to be distinguishable from O-type stars with similar
luminosity.

\section{Conclusions}\label{sect:conclusions}

We have presented atmospheric models of relatively low-mass helium stars (3.0,
4.4, and 5.1~$M_\sun$) at the final evolutionary stage, which may represent
typical SN Ib/c progenitors in binary systems. We confirm the prediction by
\citet{yoon12} that these binary progenitors can be significantly brighter than
more massive WR progenitors in the optical, and the visual brightness becomes 
higher for a less massive helium star progenitor because of 
the more extended envelope. 

Their absolute magnitudes in the optical bands are comparable to those of 20 -
30~$M_\odot$ O-type stars ($M_V  =  -5 \sim -6$ mag).  But unlike O-type stars,
their spectra could be marked by strong emission lines if the mass-loss rate is
sufficiently high.  The envelopes of helium stars of 3 - 4~$M_\sun$ undergo
rapid expansion during the late evolutionary stages to become a helium giant
($R \simeq 10 - 50~R_\sun$; \citealt{yoon10, yoon12, eldridge15}), and the
resulting emission lines would be fairly narrow, compared to those of typical
WR stars (Figures~\ref{fig2} and \ref{fig3}). 

We compared our results with the observational properties of the progenitor
candidate of the SN Ib iPTF13bvn.  We find that models with 3.0/4.4~$M_\sun$
helium star plus a 20/30~$M_\sun$ O-type star companion give the best fit
with the observation in terms of magnitudes and colors, in good agreement with
\citet{bersten14}, but a 3.0~$M_\sun$ helium star progenitor can also have
optical magnitudes comparable to those of the iPTF13bvn progenitor even without
an O-type star companion~\citep[see also][for a similar
conclusion]{eldridge15}, depending on the degree of extinction to the
source.

Future observations might find the surviving O-type star companion of the
iPTF13bvn progenitor. According to our prediction, its optical brightness
should be lower by about $\Delta M = 0.2 - 0.9$ mag than that of the observed
brightness of the iPTF13bvn progenitor, depending on the combination of the
helium star and O-type star masses.  This is because the contribution from the
helium star progenitor must have disappeared.  If no meaningful change in the
optical magnitudes of the source is found in the future, it may indicate either
that the iPTF13bvn progenitor  was more massive than about $5~M_\sun$ while the
source was the O-type companion star of $\sim 35~M_\sun$, or that the source
was not associated with iPTF13bvn at all.  A caveat in this argument is that
the luminosity of the companion star might have been significantly influenced
by the interaction with the SN ejecta, and that there even exists the
possibility that the source appears somewhat more luminous than in the
pre-SN image, because of the shock heating in the companion star~\citep{hirai15}.  
 
However, as discussed above, a $\sim 3~M_\sun$ helium star alone
without a luminous companion can also explain the optical brightness of the
iPTF13bvn progenitor in principle and non-detection of the surviving companion
star would not necessarily imply a single star progenitor because the companion
could  be a faint dwarf star or a compact object~\citep{eldridge13, yoon15}.
With reversal of the SN order in a Case A binary system, such a relatively
low-mass helium star progenitor produced in a binary system might not even have
any companion at the pre-SN stage~\citep{pols94}.

One important question still remains to be answered: if binary progenitors can
be so bright in the optical as predicted by our models and if they represent
typical SN Ib/c progenitors, why have they been so elusive in the previous
searches?  The detection limits  were $M_\mathrm{B} \simeq -4.4$ mag and
$M_\mathrm{R} \simeq -4.92$ mag in the searches of the progenitors of SN 2002ap
and SN 2010br, respectively \citep{eldridge13}. They would have been detected,
if their masses were about  $3~M_\sun$~(see Figure~\ref{fig5}), even
without a luminous companion. However, all of the other searches had more
severe detection limits ($M_\mathrm{BVRI} \lesssim -6.5$ mag), and the
progenitors would have been detected only with an O-type star companion of $M
\gtrsim 20 - 30 ~M_\sun$.  The previous difficulty in detecting a SN Ib/c
progenitor might have resulted from the fact that most SN Ib/c progenitors in
binary systems do not have a luminous O-type star companion.  
It should also be noted that  SN Ic progenitors would be generally more
difficult to detect than SN Ib progenitors. This is because SN Ib progenitors
would have a fairly massive helium envelope that results in large radii at the
pre-SN stage (hence high optical luminosities), while SN Ic progenitors
that are helium-deficient would remain very hot and faint in the
optical~\citep{yoon12}. SN 2002ap was a SN Ic, and therefore it is not
surprising that its progenitor could not be detected even with such a deep
search.  This issue should be addressed more carefully with a binary
population synthesis model that fully takes into account the evolutionary
effects on the helium star structure at the final evolutionary stage and the
recent observational constraints on the SN Ib/c ejecta
masses~\citep[cf.][]{kochanek09, yoon12, eldridge13}.

\acknowledgments
The authors are grateful to John Hillier for making the CMFGEN code publicly
available,  to John Eldridge and Melina Bersten for communicating their
results and to the anonymous referee for many useful comments that helped us improve the paper.   
S.-C. Y was  supported by the Basic Science
Research (2013R1A1A2061842) program through the National Research Foundation of
Korea (NRF).  B.-C. K was supported by Basic Science Research Program through
the National Research Foundation of Korea(NRF) funded by the Ministry of
Science, ICT and future Planning (2014R1A2A2A01002811).  H.-J. K was supported
by NRF(National Research Foundation of Korea) Grant funded by the Korean
Government (NRF-2012-Fostering Core Leaders of the Future Basic Science
Program). This work has made use of the BeSS database, operated at LESIA,
Observatoire de Meudon, France: http://basebe.obspm.fr.


\clearpage
\begin{deluxetable}{lrrrrrrrrrrrrrr}
\tabletypesize{\tiny}
\setlength{\tabcolsep}{0.07in}
\tablecaption{Model Parameters \label{tbl1}}
\tablewidth{0pt}
\tablehead{
\colhead{Model} & \colhead{$M_{\star}$} & \colhead{$L_{\star}$} & 
\colhead{$T_{\star}$\tablenotemark{a}} & 
\colhead{$T_{\rm eff}$\tablenotemark{b}} & 
\colhead{$R_{\star}$\tablenotemark{a}} & 
\colhead{$R_{\rm phot}$\tablenotemark{b}} &
\colhead{$\dot{M}$} & \colhead{$v_{\infty}$} &
\colhead{$\beta$\tablenotemark{c}} & 
\colhead{$X_{\rm H}$\tablenotemark{d}} & \colhead{$X_{\rm He}$\tablenotemark{e}} &
\colhead{$X_{\rm C}$\tablenotemark{f}} & \colhead{$X_{\rm N}$\tablenotemark{g}} &
\colhead{$X_{\rm O}$\tablenotemark{h}} \\
\colhead{ } & \colhead{($M_{\sun}$)} & \colhead{($L_{\sun}$)} & 
\colhead{(K)} & \colhead{(K)} &
\colhead{($R_{\sun}$)} & \colhead{($R_{\sun}$)} &
\colhead{($M_{\sun}$~yr$^{-1}$)} & \colhead{(km~s$^{-1}$)} &
\colhead{ } &
\colhead{ } & \colhead{ } & \colhead{ } &
\colhead{ } & \colhead{ }
}
\startdata
M3w1 &  3.0 & 4.45e+04 & 15057 & 15060 & 
30.97 & 30.97 & 6.57e-07 &  386.51 & 
1.0 & 7.26e-02 & 9.09e-01 & 1.55e-04 &
1.33e-02 & 3.71e-04 \\
M3w2 &  3.0 & 4.45e+04 & 15057 & 14990 & 
30.97 & 31.22 & 1.00e-05 &  289.88 & 
1.5 & 7.26e-02 & 9.09e-01 & 1.55e-04 &
1.33e-02 & 3.71e-04 \\
M4w1 &  4.4 & 9.05e+04 & 28596 & 28590 & 
12.26 & 12.25 & 1.50e-06 &  452.57 & 
1.0 & 1.10e-01 & 8.72e-01 & 1.22e-04 &
1.33e-02 & 4.00e-04 \\
M4w2 &  4.4 & 9.05e+04 & 28596 & 22630 & 
12.26 & 19.55 & 2.00e-05 &  339.43 & 
1.5 & 1.10e-01 & 8.72e-01 & 1.22e-04 &
1.33e-02 & 4.00e-04 \\
M5w1 &  5.1 & 1.12e+05 & 50587 & 49520 & 
 4.35 &  4.54 & 4.88e-06 & 1332.76 & 
1.0 & 7.35e-06 & 9.81e-01 & 2.46e-04 &
1.32e-02 & 2.94e-04 \\
M5w2 &  5.1 & 1.12e+05 & 50587 & 39430 & 
 4.35 &  7.16 & 2.00e-05 &  999.57 & 
1.5 & 7.35e-06 & 9.81e-01 & 2.46e-04 &
1.32e-02 & 2.94e-04 \\
Ostar1 & 20.0 & 7.05e+04 & 29402 & 29400 & 
10.23 & 10.22 & 1.50e-07 & 1714.39 & 
1.0 & 7.01e-01 & 2.80e-01 & 3.48e-03 &
1.03e-03 & 1.00e-02 \\
Ostar2 & 30.0 & 1.99e+05 & 32088 & 32090 & 
15.00 & 14.41 & 6.00e-07 & 1718.73 & 
1.0 & 7.01e-01 & 2.80e-01 & 3.48e-03 &
1.03e-03 & 1.00e-02 \\
Ostar3 & 35.0 & 3.72e+05 & 28610 & 28610 & 
24.84 & 24.82 & 1.13e-06 & 1458.47 & 
1.0 & 6.46e-01 & 3.36e-01 & 1.99e-03 &
3.99e-03 &  8.29e-03 \\
\enddata
\tablecomments{ All the parameters except $T_{\rm eff}$ and $R_*$
are taken from the stellar evolueionary models of \citet{yoon10} and used 
as the input parameters for the CMFGEN calculations. 
$T_{\rm eff}$ and $R_{\rm phot}$ are the outputs from the calculations.}
\tablenotetext{a}{From the stellar evolution code \citep{yoon10} 
without correcting the optical depth effects from the wind.}
\tablenotetext{b}{Effective temperature and photospheric 
radius from the CMFGEN calculations defined 
at the Rosseland optical depth = 2/3.}
\tablenotetext{c}{From the standard $\beta$-type velocity 
law. $\beta$=1 and 1.5 for w1 and w2 models, respectively.}
\tablenotetext{d,e,f,g,h}{Mass fraction of hydrogen, helium, carbon, 
nitrogen, and oxygen.}
\end{deluxetable}

\begin{deluxetable}{lcc}
\tabletypesize{\scriptsize}
\tablecaption{Model Atoms Used in CMFGEN Calculation \label{tbl2}}
\tablewidth{0pt}
\tablehead{
\colhead{Species} & \colhead{Super Levels} &
\colhead{Full Levels}
}
\startdata
H~\textsc{i} & 20 & 30 \\
He~\textsc{i} & 45 & 69 \\
He~\textsc{ii} & 22 & 30 \\
C~\textsc{ii} & 40 & 92 \\
C~\textsc{iii} & 51 & 84 \\
C~\textsc{iv} & 59 & 64 \\
N~\textsc{ii} & 45 & 85 \\
N~\textsc{iii} & 41 & 82 \\
N~\textsc{iv} & 44 & 76 \\
N~\textsc{v} & 41 & 49 \\
O~\textsc{ii} & 54 & 123 \\
O~\textsc{iii} & 88 & 170 \\
O~\textsc{iv} & 38 & 78 \\
O~\textsc{v} & 32 & 56 \\
O~\textsc{vi} & 25 & 31 \\
Si~\textsc{iii} & 33 & 33 \\
Si~\textsc{iv} & 22 & 33 \\
Fe~\textsc{iii} & 104 & 1433 \\
Fe~\textsc{iv} & 74 & 540 \\
Fe~\textsc{v} & 50 & 220 \\
Fe~\textsc{vi} & 44 & 433 \\
Fe~\textsc{vii} & 29 & 153 \\
\enddata
\tablecomments{
The second column denotes the total number of super levels, which mean the
energy levels that was used for the calculation.  In the calculation, some
levels having similar excitation energies are grouped into a single super
level, to save computing time.  In the case of H~\textsc{i}, for example, 30
levels are considered in the calculation.  For the first 15 levels, each super
level corresponds to the actual energy level of  H~\textsc{i}, but the super
levels from 16 to 20 correspond to the grouped levels for 16--18, 19--21,
22-24, 25--27, and 28--30, respectively.  Therefore, the total number of super
levels used for the computation is 20, while the total number of the considered
energy levels is 30, which is given in the third column. 
} \end{deluxetable}

\begin{deluxetable}{lcccccc}
\tabletypesize{\scriptsize}
\tablecaption{Absolute magnitudes and bolometric corrections 
of the models and the blackbody approximations 
in the {\it HST}/ACS {\it F435W}, {\it F555W}, 
and {\it F814W} filters. \label{tbl3}}
\tablewidth{0pt}
\tablehead{
\colhead{Model} & \colhead{$M_{F435W}$} &
\colhead{$M_{F555W}$} & \colhead{$M_{F814W}$} &
\colhead{BC$_{F435W}$} & \colhead{BC$_{F555W}$} &
\colhead{BC$_{F814W}$} \\
\colhead{ } & \colhead{(mag)} &
\colhead{(mag)} & \colhead{(mag)} &
\colhead{(mag)} & \colhead{(mag)} &
\colhead{(mag)} 
}
\startdata
M3w1 & -5.87 & -5.71 & -5.54 & -1.01 & -1.17 & -1.34 \\
M3w2 & -5.85 & -5.73 & -5.65 & -1.03 & -1.15 & -1.23 \\
M4w1 & -5.21 & -4.98 & -4.68 & -2.44 & -2.67 & -2.97 \\
M4w2 & -5.53 & -5.42 & -5.30 & -2.12 & -2.23 & -2.35 \\
M5w1 & -4.13 & -3.79 & -3.63 & -3.75 & -4.09 & -4.25 \\
M5w2 & -4.83 & -4.57 & -4.54 & -3.05 & -3.31 & -3.34 \\
Ostar1 & -4.86 & -4.61 & -4.28 & -2.52 & -2.77 & -3.10 \\
Ostar2 & -5.79 & -5.54 & -5.22 & -2.72 & -2.97 & -3.29 \\
Ostar3 & -6.75 & -6.52 & -6.22 & -2.44 & -2.67 & -2.97 \\
\tableline
\multicolumn{7}{c}{In binary systems (from composite spectra)} \\
\tableline
M3w1+Ostar1 & -6.23 & -6.05 & -5.83 & \nodata & \nodata & \nodata \\
M3w2+Ostar1 & -6.22 & -6.06 & -5.92 & \nodata & \nodata & \nodata \\
M4w1+Ostar1 & -5.80 & -5.56 & -5.25 & \nodata & \nodata & \nodata \\
M4w2+Ostar1 & -6.00 & -5.84 & -5.66 & \nodata & \nodata & \nodata \\
M5w1+Ostar1 & -5.31 & -5.03 & -4.75 & \nodata & \nodata & \nodata \\
M5w2+Ostar1 & -5.60 & -5.34 & -5.17 & \nodata & \nodata & \nodata \\
M3w1+Ostar2 & -6.58 & -6.38 & -6.14 & \nodata & \nodata & \nodata \\
M3w2+Ostar2 & -6.57 & -6.39 & -6.20 & \nodata & \nodata & \nodata \\
M4w1+Ostar2 & -6.29 & -6.05 & -5.73 & \nodata & \nodata & \nodata \\
M4w2+Ostar2 & -6.42 & -6.23 & -6.01 & \nodata & \nodata & \nodata \\
M5w1+Ostar2 & -6.00 & -5.74 & -5.44 & \nodata & \nodata & \nodata \\
M5w2+Ostar2 & -6.16 & -5.91 & -5.68 & \nodata & \nodata & \nodata \\
M3w1+Ostar3 & -7.15 & -6.94 & -6.68 & \nodata & \nodata & \nodata \\
M3w2+Ostar3 & -7.14 & -6.95 & -6.72 & \nodata & \nodata & \nodata \\
M4w1+Ostar3 & -6.98 & -6.76 & -6.45 & \nodata & \nodata & \nodata \\
M4w2+Ostar3 & -7.05 & -6.86 & -6.60 & \nodata & \nodata & \nodata \\
M5w1+Ostar3 & -6.84 & -6.60 & -6.31 & \nodata & \nodata & \nodata \\
M5w2+Ostar3 & -6.92 & -6.69 & -6.43 & \nodata & \nodata & \nodata \\
\tableline
\multicolumn{7}{c}{Blackbody\tablenotemark{a}} \\
\tableline
BB\_M3 & -5.91 & -5.85 & -5.81 & -0.97 & -1.03 & -1.07 \\
BB\_M4 & -5.32 & -5.10 & -4.85 & -2.33 & -2.55 & -2.80 \\
BB\_M5 & -4.01 & -3.72 & -3.38 & -3.87 & -4.16 & -4.50 \\
\enddata
\tablenotetext{a}{Based on the blackbody fluxes for the given 
temperature ($T_{\star}$) and luminosity of 3.0, 4.4, and 5.1~$M_{\sun}$ 
models in Table~\ref{tbl1}.}
\end{deluxetable}

\end{document}